# Space ethics to test directed panspermia*

Maxim A. Makukov[a] and Vladimir I. *sh*Cherbak[b]

The hypothesis that Earth was intentionally seeded with life by a preceding extraterrestrial civilization is believed to be currently untestable. However, analysis of the situation where humans themselves embark on seeding other planetary systems motivated by survival and propagation of life reveals at least two ethical issues calling for specific solutions. Assuming that generally intelligence evolves ethically as it evolves technologically, the same considerations might be applied to test the hypothesis of directed panspermia: if life on Earth was seeded intentionally, the two ethical requirements are expected to be satisfied, what appears to be the case.

## 1. Introduction

The emergence of life, whichever mechanisms stand behind it, is apparently a rare event. This is a fairly safe presupposition, as it falls into neither of the two extremes. On one extreme, it is supposed that life emerges obligatorily whenever necessary conditions occur[1,2], and that such conditions might be common in the Universe. On the other extreme, abiogenesis is thought to involve a complex series of accidental events improbable to the extent that it requires recourse to an infinite multiverse[3], or even considered to be unknowable in principle[4]. In between is the presupposition that abiogenesis can occur comprehensibly in a finite universe, though it involves accidental events or requires rarely occurring set of circumstances, or both, making it a rare phenomenon. The failure to easily simulate abiogenesis in a variety of conducive conditions[5] seems to support the conclusion that the required set of circumstances is specific and therefore probably uncommon.

However, the nature of life is such that, once started, it tends to reproduce as much as possible, colonizing even harsh environments as exemplified by all kinds of extremophiles. This inherent tendency of life to reproduce is the very prerequisite for biological evolution, making it possible for complex features facilitating further propagation to emerge. The most powerful of such features is intelligence[6]. Without it, the inherent expansion of life is limited to a single planet. With it, life might propagate throughout the Galaxy or even entire Universe.

There are two possible modes of life's cosmic expansion at intelligent stage: intelligent beings might colonize other planetary systems themselves or they might implant them with microbial life. The first mode (interstellar colonization[7]) propagates life and intelligence simultaneously, but is highly demanding and hazardous. The second mode (intentional seeding, or directed panspermia[8]) propagates as yet unintelligent life but is far more feasible technologically. As for natural interstellar panspermia, it appears to be overwhelmingly unlikely[9,10], or at best uncertain[11,12], as there are many "ifs" related to both the incidental nature of the process and viability of microorganisms under a number of detrimental conditions. In intentional seeding all these factors are obviated or alleviated, making it much more efficient for dispersal of life within the Galaxy. Starting local evolution, some of the seeds that happen to inhabit a planet with suitable conditions might eventually arrive at intelligent stage[13,14] and repeat the cycle of cosmic expansion.

The efficiency of intentional seeding, together with the data indicating that first habitable planets were present long before the Solar System formed[15,16], suggests that at the current age of the Galaxy it might be even more probable for an intelligent being to find itself on a planet where life resulted from directed panspermia rather than on a planet where local abiogenesis took place, and the Earth is not an exception from that. This is not to say that the view that terrestrial life originated locally is flawed. But subscribing largely to this view and dismissing the possibility that terrestrial life might not be a first independent generation in the Galaxy is probably nothing but a manifestation of geo-anthropo-centrism (inappropriately armed with Occam's razor[17]). Particularly so in view of the universality of the terrestrial genetic code which counts more in favor of seeding rather than of local abiogenesis (though it also might count in favor of non-directed panspermia).

The hypothesis that terrestrial life derives from intentional seeding by an earlier galactic civilization is far from being new. It was first touched upon in the middle of the 20th century by John Haldane[18] and considered later by Carl Sagan[19] and, in great detail, by Francis Crick[20,21] and Leslie Orgel[20]. The typical arguments raised against this hypothesis are that it is untestable and that it leaves the issue of motivations for seeding open (yet, the former argument is equally applicable to the hypothesis of local abiogenesis which does not preclude it from being the default view in astrobiology[22]). However, here we show that, in fact, intentional seeding might have at least two observable aspects. Though neither of the two is an indispensable concomitant of seeding, one of them, if present, serves as the *experimentum crucis* and, furthermore, resolves the issue of motivations. To show how these aspects arise, we apply the same principle of detailed cosmic reversibility that Crick & Orgel[20] used to demonstrate the very plausibility of the hypothesis. In other words, we first envisage the situation where humans themselves embark on seeding other planetary systems.

## 2. Space ethics

According to the online poll of the Space Settlement Institute[23], the major argument for space colonization, more important than gaining new energy resources, is ensuring survival of humanity and terrestrial life in general from global calamities that might happen on Earth (see also Refs. 24-26). On larger scales the same argument is applicable to the entire Solar System[27,28]. However, interstellar colonization, if feasible at all, is technologically far more uncertain than near-solar settlements[29,30]. But unmanned interstellar missions are already a reality[31], and are likely to be improved in the relatively near future[32]. This makes it possible to secure terrestrial life by seeding other planetary systems, as already has been proposed[8,33-35] (planetary seeding within the Solar System was considered even earlier[36-38]). Shielded from radiation in automated probes and deep-frozen in a suitable media, bacterial spores and microorganisms capable of cryptobiosis[39] might easily survive the time span of the interstellar travel, which, with proper propulsion technology, is substantially reduced compared to natural panspermia. On arrival at target exoplanets, recovering metabolic activity and adapting to local environments, the seeds are expected to trigger off local evolution. The chances of successful seeding might be increased with sending different types of seeds[8,21], including varieties of extremophiles both of chemo- and photoautotrophic types, as well as eu-karyotes to facilitate higher evolution. Besides, bioengineering might help to improve adaptive capabilities of the seeds and, if known, adjust them to the specifics of the target environment[8].

However, cosmic seeding raises two issues concerned with ethics (it should be noted that the above-mentioned argument for the expansion of life in space implies valuing the phenomenon of life and thus itself has biotic ethics in its core[40]; note also that this motivation for seeding is especially justified if the presupposition of rare abiogenesis is valid). The first issue, relevant in colonization as well, goes back to the UN Outer Space Treaty[41] and is related to the general subject of planetary protection[42]. Given that the purpose of seeding (protecting life itself) stems from intrinsic value, this activity does not appear to contradict those principles of the Outer Space Treaty and follow-up regulations which stem from instrumental and aesthetic values. But the problem shows up if some of the target planets are already inhabited with indigenous lifeforms[21,33,40,43-45]. In general, norms of treating extraterrestrial life depend on both the accepted ethical system[43] (e.g., eco- or anthropocentrism) and whether the life in question is microbial, sentient or intelligent[44]. However, the purpose of seeding is such that it might comply with all (or at least most[46]) ethical systems by avoiding potentially inhabited planets. The necessary requirement therefore for the target exoplanets is that they must not reveal any kind of biosignatures[47]. As this requirement is obviously insufficient, the most straightforward way to further minimize the risk of interfering with indigenous life is to select newborn exoplanets[40] or even planetary systems still at the stage of formation. However, as the majority of stars are formed in a clustered mode[48,49], the most efficient strategy proves to be in seeding collapsing clouds that form open star clusters[50]. Not only that minimizes the risk of interfering, but natural panspermia within clusters[10,51] (which is far more likely than natural transfer of viable cells between field stars) and within infant planetary systems[52] might greatly amplify the result. Many seeded planets are then dispersed throughout the Galaxy with their host stars as clusters quickly dissolve in the galactic disc within few millions of years after formation[49]. Besides, seeding star-forming clouds does not require sophisticated technologies for extremely precise navigation needed to seed individual exoplanets.

The second ethical issue is seeding-specific and is even more acute as it is concerned with potentially intelligent life. In cosmic seeding the tacit hope is that the seeds will eventually evolve into intelligent forms (in fact, given proper conditions this might be not unlikely as suggested by evolutionary convergencies[13,14]) and achieve scientifically advanced stage. If that happens, however, the question arises if it is ethical to leave those intelligent beings unaware of the fact that life on their planet derives from seeding activity by a preceding civilization. This question is no more in the realm of environmental ethics, but rather in the realm of moral obligations to rational beings[53]. Assuming that this question is negatively answered (at least because that will save their efforts in explaining spontaneous emergence of life specifically in local environments), we should seek to provide our cosmic descendants with at least minimum knowledge that life on their planet derives from seeding by intelligent predecessors. An accompanying artifact made of any passive material and carrying a message is of no use here as it will erode long before intelligent beings evolve (even if it endures there is no guarantee it will be found). As a suitable and perhaps the only workaround, the message might be put inside the seeds themselves. DNA is already used to store non-biological information[54-57], so embedding a message into genomes is not a problem (in fact, genomes were considered as possible carriers for interstellar messages[54,58-62]). The problem is that we do not know how to prevent that message from being destroyed by mutations during evolution. We need something that goes unchanged with the cells as they replicate for billions of years and yet might be modified artificially. There is only one thing in the cell that meets this requirement: the genetic code.

The genetic code is extremely stable through time due to strong purifying selection, as correct translation of all genes in genomes hinges on it. The mapping of the code is exactly the same in practically all organisms on Earth, implying that it has been unchanged since the last universal common ancestor (LUCA), i.e. for almost four billion years[63]. Only under specific conditions, especially in small genomes, codon reassignments might become fixed[64,65], and there are a couple of dozen known[66,67] (and probably more still unknown[68]) variations of the code in simple organisms and organelles. Nevertheless, the fact that the vast majority of even simple organisms still use LUCA's code suggests that it might serve as an exceptionally reliable storage for a small message. The question then comes to if it is technically possible to insert a message into the code and, if yes, what kind of message that could be given the specifics of the carrier.

## 3. The message that goes with its addressee

Inserting any kind of message into the code implies its modification. As learnt from past decades, the code is quite amenable to artificial codon reassignment[69,70], as well as increasing[71,72] and decreasing[73] the number of encoded amino acids and even altering the length of codons[74]. In principle, there seems to be no fundamental barrier to reassigning all or at least most of the 64 codons in a living cell provided that all genes are appropriately rewritten to leave the encoded proteins unaltered (rewriting of genes is also possible when entire genomes are synthesized from scratch[55]). Admittedly, there might be certain technical challenges – after all, the main function of the code is to efficiently translate genes, while the message might be only piggy-backed on that. For example, the genetic code is known to be near-optimal at minimizing the effect of mutations and misreadings[75,76]. This feature, however, might be taken as a requirement in constructing the message, formalized with the same error cost function[76] used to estimate the robustness of the code. Another complication is that the genetic code overlaps with a variety of non-translational mechanisms[77-79]. Thus, through synonymous codon usage the redundancy pattern of the code affects expression[80] and folding[81] of proteins and binding of transcription factors[82]. But as insertion of the message should leave both the amino acid repertoire and the average redundancy pattern unchanged (as might be required by the efficiency of codon-anticodon recognition at the ribosome[83,84]), those effects, in principle, might be restored with the new code as well (gene expression might also be readjusted at other levels, e.g., by tuning transcriptional control). Besides, a recent study[85] suggests that genome-wide recoding of genes is feasible even without special measures to compensate for all those interconnections, at least in prokaryotes.

We are left then with the question as to what kind of message might be inserted into the genetic code given its general character and functional constraints. This is where the experience of messaging to extraterrestrial intelligence[86-89] (METI) might come to help. In case of genomes it is straightforward to insert a bitmap pictogram[54] conceptually analogous to radio pictograms sent from the Earth[89]. However, in case of the genetic code the message cannot be of a pictorial type, since elements of the carrier (codon–amino acid pairs) do not form spatial (like nucleotides in DNA) or temporal (like radio beeps) sequence. But this is not a drawback since the concept of pictorial representations relies on vision[90] whereas ideally the message should not be reliant on any particular sensory modality[91]. Next, if cognitive universals do exist, mathematics and logic are believed to be the first candidates for that[92-95]. Mathematics might be introduced in case of the genetic code by quantifying certain properties of amino acids. Parameters chosen for quantification must not rely on conventional systems of units and therefore should be of simple countable type. Atomic or nucleon numbers of amino acid molecules seem to be the simplest candidates for that. Indeed, the number of certain objects (e.g., nucleons in a molecule) relies neither on particular sensory modality nor on systems of units. The final step is to construct the message using the chosen parameter. It cannot be the series of primes or Fibonacci numbers[87] (traditional favorites in METI) since here we are restricted with the fixed set of numbers representing the amino acids. What we can do is project such a mapping between amino acids and codons which conforms to functional requirements and, at the same time, reveals a special feature in its mathematical structure. This might be, e.g., a solved puzzle of some kind, conceptually similar to Sudoku[96], magic squares[97] and other combinatorial puzzles. In fact, sending solved games and puzzles is not a new concept in METI[98], and the nature of the genetic code optimally fits this type of messaging.

Another advantage of the "genetic METI" over external artifacts is that it guarantees that the message will be discovered no sooner than the recipient becomes capable of cracking the genetic code, i.e. achieves scientifically advanced stage. That precludes interference with the pre-scientific cultural evolution – another ethical issue by itself, related in its character to the ethics-based version[99] of the zoo hypothesis[100] (anticipated by Tsiolkovsky at the dawn of the cosmic era[101]).

## 4. Detailed cosmic reversibility: The case of terrestrial life

As there is no compelling evidence that terrestrial life originated locally on Earth, the possibility that it was seeded by an earlier galactic civilization[18-21] cannot be excluded. The very ability to perform intentional seeding implies advanced technological stage, which, in its turn, is hardly achievable without developing ethically[87,102]. Therefore, it is safe to assume that any civilization capable of intentional seeding would take care of the ethical issues, so the above ethical reasoning might be applied to test the hypothesis of directed panspermia.

The first ethical requirement of non-interfering is immediately found to be satisfied in this case: it is well established that life was flourishing already early in Earth's history[103-105], and might have been present at final stages of Earth formation[106] and earlier, when the Sun was still a member of a star cluster[10,107]. But this fact might also be interpreted as an indication of the rapidity of abiogenesis[108,109] (on Earth or anywhere in the original cluster), so it represents supportive but certainly insufficient evidence for directed panspermia.

Turning to the second ethical requirement we should analyze the standard version of the terrestrial genetic code for some kind of intelligent hallmark (this possible concomitant of directed panspermia which links the problem of the origin of life on Earth to the problem of searching for extraterrestrial intelligence was first pointed out by György Marx[58,60]). Before that, however, it is appropriate to review data that could potentially argue against the possibility of a message in the terrestrial code.

First of all, there are several distinct models seeking to provide rationale for the mapping of the standard code[110-113] (with the major ones being adaptive, stereochemical and biosynthetic models), but the general conclusion is that the question is still open[111,113]. Thus, rough correlation between first bases of codons and biosynthetic pathways of corresponding amino acids[114] seems to support the biosynthetic model. However, statistical significance of this pattern does not appear to be extremely high, and the adequacy of the model itself has been questioned[115]. The most discussed feature of the code – its robustness to errors attributed to either adaptive[116] or neutral[117] evolution – cannot preclude the possibility of a message: as discussed above, it might have been introduced (together with the long-term impact of the code architecture on evolutionary dynamics[118,119]) as a formalized requirement in constructing the message-harboring mapping. The adaptiveness of the amino acid alphabet itself[120] is unrelated to the mapping and so is not indicative in this regard. RNA-binding data for some of the amino acids seems to point at stereochemical rationale for the code[121]. Apart from uncertainty in its relevance to the actual code origin[122] and lack of independent verification, this data also cannot preclude the possibility of a message since its insertion might have not required reassignment of all codons (in fact, this would be even desirable for simplification of the embedding procedure and minimization of possible adverse effects), leaving a stereochemical core from the primordial code. Another study based on comparative tRNA analysis suggests that "the genetic code is not older than, but almost as old as our planet"[123]. This follows from the conclusion that "had the code been much older – and this would be possible only in case of extraterrestrial origin – those changes that clearly can be identified as phylogenetic divergence would previously have become randomized to a large extent". However, it is not obvious that low phylogenetic divergence should unambiguously indicate small period elapsed since the emergence of life. It might as well indicate a bottleneck event in between (and in effect, panspermia, directed or not, amounts to such an event), and that the sequences in question evolve under stringent constraint (not surprising as translational machinery is indeed the most conserved element of cellular life[63]). (Ironically, using genomic analysis other researchers come up with the opposite conclusion that terrestrial life is at least 7-8 billion years old[124], thereby ruling out both local abiogenesis and intentional seeding, and leaving natural panspermia as the only viable option; however, see Ref. 125).

The rest of potential objections boil down to the following one: there is so much more to the genetic code than just a mapping that it is unreasonable to expect a message there. Indeed, there are non-standard coding schemes[126], involvement of both tRNAs[127] and synthetases[128] in non-translational processes in higher organisms, effect of the redundancy pattern of the code (through synonymous codon usage[78,129]) on expression[80], folding[81] and structure[130] of proteins, translation accuracy[131], cell cycle regulation[132], nucleosome positioning[133], splicing[134], secondary mRNA structure[135], and binding of transcription factors[82]. However, these objections are irrelevant since all (or at least most of) these features are the result of post-LUCA (i.e. post-seeding) evolution and adaptation of certain mechanisms to the redundancy pattern of the code, not the other way round. This follows from the fact that whereas the genetic code is universal (its variations also represent post-LUCA modifications[64,66]), these features are not, either missing in some domains or differing in details between and within domains. The same goes for the objection that if global codon reassignment was performed in LUCA (which is nothing but the original seeds in case of directed panspermia) for inserting a message, we would expect that today artificial global reassignment of codons would also be quite easy; yet, today it is challenging, particularly because in most tRNAs anticodons are involved in the recognition by synthetases, making the genetic code essentially "hardwired" for its most part. But the very fact that details of the structure, recognition and aminoacylation of homologous tRNAs differ in various organisms[136] points at their post-LUCA evolution, and it might be that in LUCA the code was much less hardwired (in fact, freedom in reassignment at pre-LUCA times is required by the dominant adaptive model of the code evolution). To conclude, testing the standard code for a message does not appear preposterous.

The result of such test, with amino acid nucleon number as the messaging parameter, suggests that the second ethical requirement is also satisfied. Referring to the original paper[137] for the detailed description, here we outline only essential points. Apart from its biological function, the standard version of the terrestrial code appears to piggy-back precision-type mathematical and ideographical structures which are unexpected in traditional models of the code evolution but make good sense within the hypothesis of directed panspermia. Both mathematical and ideographical parts of what appears to be the message share common style and are statistically profound to the extent that the code mapping itself might be uniquely deduced from them.

The mathematical part might be described as a solved puzzle with the rules formulated as follows: preserving the redundancy pattern on average uniform and error cost low, find a mapping between triplets and amino acids that reveals precise equalities of nucleon sums in as many logical arrangements of the code as possible, accompanied, whenever possible, by distinctive notation of those sums in one and the same positional numeral system and by at least one of the three possible pair inversions between nucleotides. The solution to this puzzle implemented in the standard mapping of the code comprises eleven algebraically independent equalities of nucleon sums in seven simple logical arrangements of the code; almost all of these equalities display distinctive notation in the decimal system, and almost all of the arrangements involve at least one of the pair inversions between nucleotides. This is quite impressive, as even without ideographical part and functional constraints achieving a similar result is a computationally nontrivial combinatorial task: similar to Sudoku where each element participates in three overlapping patterns, each amino acid contributes its nucleons simultaneously to seven or more overlapping nucleon equalities. Analysis of other known versions of the code did not reveal any arithmetical structures similar in complexity to that found in the standard code, in agreement with that these versions represent post-LUCA variations.

Symmetrization of the standard code produces the mapping differing only in one spot (with TGA-codon reassigned from stop-signal to cysteine). Even with this reassignment of a single codon, the symmetrized version no longer contains the entire puzzle, but straightforward alignment of its codons still using nucleon numbers of their amino acids leads to a multilevel ideographical structure displaying a bunch of related symbolic and syntactic symmetries. This whole structure is revealed provided that zero (represented by stop-codons) is positioned properly in front of its series. That implies that zero, apart from being implicitly used in positional notation in the mathematical part, is also displayed explicitly as an individual symbol in the ideographical part. By themselves, both the preferred numeral system[138] and the symbol of zero[139,140] are strong hallmarks of intelligence. On the whole, the message is of the type that "requires reconstruction, but designed to facilitate decoding"[91] (through universal nucleon transfer in proline for the mathematical structure[137] and through symmetrization of the code for the ideographical part). Facilitated decoding implies involvement of recipient's intelligence into the process of reconstruction and thus represents a good strategy to protect the message from being misconstrued with natural causes (the problem of SETI in general, particularly so in case of biological media).

The processes of molecular evolution are stochastic (even if under non-random forces) and free of interpretive semiosis in their nature. Therefore, unlike the first case of life's early appearance on Earth, statistically strong patterns in the genetic code which reveal both punctually precise character and symbolic representations do not seem to allow for ambiguous construal, and thus represent sufficient evidence for directed panspermia.

## 5. Implications for evolutionary biology and astrobiology

Rooting the tree of life, i.e. locating LUCA's position relative to the major stems of the phylogenetic tree, is a notorious issue in evolutionary biology[141,142]. Whether LUCA emerged locally from protobionts or was brought here as the seeding material might affect conclusions about its character. Thus, if the original seeds were represented by cells of various types, the tree of terrestrial life has in fact multiple (though not independent) roots[143]. Furthermore, as some eukaryotes are especially good at both cryptobiosis[39] and adaptability[144,145], it might be that the original seeds were of eukaryote-like type[21], with prokaryotes emerging from them through reductive evolution – the possibility suggested by other considerations as well[146,147]. Another possible evidence for multiple seed types is that the symmetrized code version, which carries the ideographical part of the message, is found in euplotid ciliates[137] (which are eukaryotes). However, a careful comparative study is needed to determine if the euplotid version might be as old as the standard code, or if it represents a later coincidence (which is not improbable given that TGA is one of the spots in the code that have been reassigned independently during evolution in various lineages, presumably in transition to higher thermodynamic stability[66]).

The implication for astrobiology concerns two parameters in the Drake equation. First, in its original form the equation assumes that life emerges independently in individual planetary systems (this assumption holds also for the Rio scale[148] developed to assess the significance of putative SETI signals and artifacts, so it cannot be applied adequately to the artifact in the genetic code). It was later suggested to update the Drake equation taking into account interstellar colonization[149] and seeding[150]. The extent of seeding by a single civilization is limited only by economical factors, so it might increase the $f_l$-term in the equation (fraction of habitable planets inhabited with simple life) by orders of magnitude, especially with seeding cluster-forming clouds. Second, $f_i$-parameter describing the probability of intelligence arising from simple life-forms has been considered completely uncertain, ranging from near zero, with intelligence being an evolutionary fluke[151-154], to near one, with intelligence being a quite probable product of convergent evolution[6,13,14,155-157]. The message in the genetic code suggests that, apart from our Solar System, intelligence had emerged elsewhere at least once, implying that $f_i$-term is not near zero after all (under the right set of astrophysical and planetary circumstances[153]). Thus, even if abiogenesis occurred only once in the Galaxy, the outcome of the Drake equation might be greater than one. This does not lead to the Fermi paradox[158] which is in fact naturally resolved in this case: the Earth *was* colonized, via simple life-forms[159].

The presence of the message in the genetic code also resolves the issue of motivations in directed panspermia. Thus, it was suggested that possible motivations for seeding might include long-term evolution experiment and terraforming for future colonization[19,159]. In these cases some kind of a hallmark in the seeds is desirable indeed, e.g., for subsequent contamination control. However, insertion of a complex signature displaying notions of zero, positional notation and other semiotic features is too disproportionate for those purposes. As such signature is evidently designed to be universally intelligible, the only plausible motivation appears to be the propagation of life and intelligence in the Universe. This brings us to another implication: if our civilization itself ever comes to cosmic seeding, it will not have to be concerned about embedding a message into the seeds. With triplet code it is hardly possible to compose a family emblem more elaborate than the one which is already in place.